\UseRawInputEncoding 
\documentclass{resonance}

\begin{document}

\title{Will Betelguese Explode?}
\secondTitle{}
\author{Priya Hasan}

\maketitle
\authorIntro{\includegraphics[width=2cm]{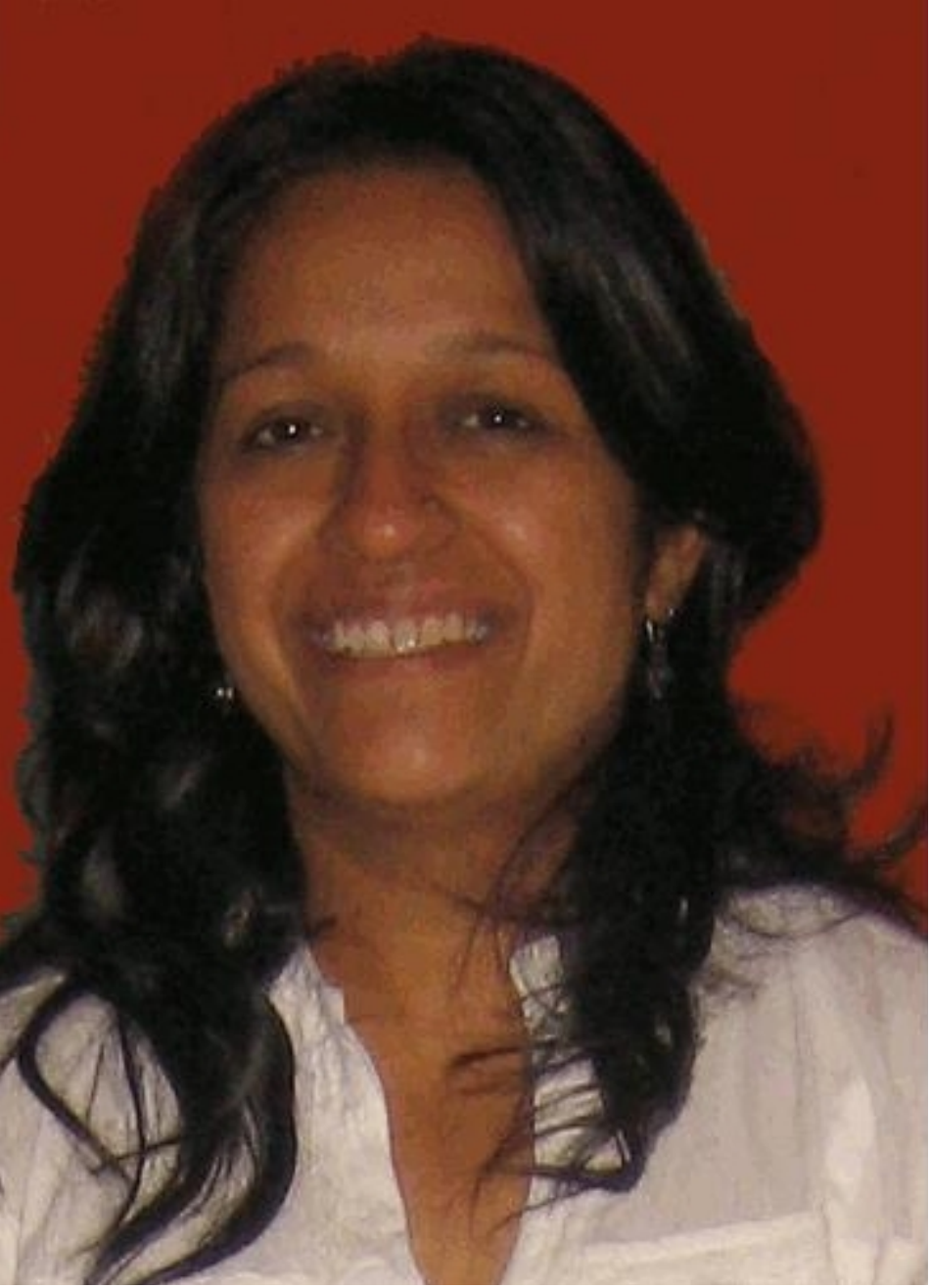}\\
Dr Priya Hasan is an Asst~Professor in Physics at the Maulana Azad National Urdu University, Hyderabad. Her research interests are observational astronomy, star formation, star clusters and galaxies.  She is actively involved in olympiads, public outreach and astronomy education programs. She is the Secretary of the Public  Outreach \& Education Committee of the  Astronomical Society of India.}
\begin{abstract}
Since October 2019, Betelgeuse began to dim noticeably and by January 2020 its brightness had dropped by a factor of approximately 2.5, demoting it from the position of the top (apparent) brightest 11 th star to the 21 st!!! 
Astronomers were excited and thought of it as the lull before the storm, Betelguese was ready to go supernova!!!

 This article is aimed more as a case study where we show how this question was answered using scientific arguments and data. It will also highlight the importance of supernovae to human existence and give a brief discussion on the evolution of massive stars.  
 
 And also, answer the question!!!
\end{abstract}

\monthyear{April 2020}
\artNature{GENERAL  ARTICLE}

\begin{center}

\it{`Without these supernova explosions, there are no mist-covered swamps, computer chips, trilobites, Mozart or the tears of a little girl. Without exploding stars, perhaps there could be a heaven, but there is certainly no Earth.' 
  \\
  Clifford A. Pickover}
 \end{center}

\section*{Introduction}
The answer to the question `Will Betelguese Explode?' is obviously Yes! But when is the real question.

Spiral galaxies such as the Milky Way are supposed to host about three supernovae per century. We thus expect to see as many as 60 supernova explosions that are younger than 2,000 years old, but fewer than 10 have been found. The most recent supernova to be seen in the Milky Way galaxy was SN 1604 also called `Kepler's star' which was visible with the naked eye on October~9, 1604 in the constellation Ophiuchus. A supernova remnant within our own galaxy in the constellation Cassiopeia, called Cassiopeia A, occurred in 1680, but  was  discovered centuries later in 1948 by Martin Ryle and Francis Graham-Smith, astronomers at Cambridge, based on observations with the Long Michelson Interferometer. The optical component was first identified in 1950. Cas~A is 3C461 in the Third Cambridge Catalogue of Radio Sources and G111. 
\leftHighlight{The constellation of Orion, the Hunter, appears in the winter sky, with his bow and his hunting dogs, Canis Major and Canis Minor, close to his feet. Orion is located on the celestial equator, is one of the most prominent and recognizable constellations in the winter sky of the northern hemisphere.}
 An more recent supernova remnant was discovered in 1984 close to the galactic center in the radio with the Very Large Array (VLA). It is perhaps the youngest of all supernovae in the Milky Way and is as young as 100--200 yrs old. No other supernova that has occurred in our galaxy has been observed since.
\begin{figure}[!t]
\caption{Betelguese: The right shoulder of Orion. (Image Credit:  Wikipedia)}
\label{betel} 
\vskip -12pt
\centering
\includegraphics[width=5.5cm, height=7.0cm]{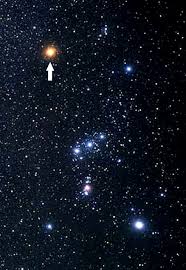}
\end{figure}

$\alpha$ \textit{Orionis} or  Beteleguese is a very well known, easy to identify star in the constellation Orion (Fig \ref{betel}). Since October 2019, it began to dim noticeably, and by January 2020 its brightness was reduced 2.5 times, from magnitude 0.5 to 1.5. Astronomers all over the world were excited that this was probably the dimming due to contraction of the star which would be followed by a supernova explosion. There were various articles on how it would be like witnessing such a magnificent event. And the excitement continues. 

In this article an attempt has been made to explain this possibility, its implications and the method in which investigations were made to answer this question.

\section{Betelguese: Fast Facts}

Lets first gather our facts about Betelguese. It is a red super giant (RSG) at a distance of $222\pm40$ pc  with a progenitor  main sequence mass of $20^{+5}_{-3}$ M$_{\odot}$. It has an age of $8-8.5$ Myr \cite{dolan16, leves20} and is  in the red giant phase of evolution  and hence it has expanded to a size of $887\pm203$ R$_{\odot}$, however it has cooled to a temperature of $\approx$3600 K.

\rightHighlight{\textbf{Fast Facts on Betelguese:}\\ Distance:  222$\pm$40 pc \\ Age: 8--8.5~Myr\\Progenitor Main Sequence Mass: $20^{+5}_{-3} $M$_{\odot}$ \\ Radius $887\pm203$ R$_{\odot}$ \\ Rotation:  8.4 years\\ Luminosity: 125,000  L$_{\odot}$ \\ Mass loss: $10^{-6} $ M$_{\odot}$ yr$^{-1}$}

Due to expansion and hence conservation of angular momentum, giants tend to rotate slowly.  Betelgeuse  spins  once every 8.4 years, whereas the Sun spins once a month. Luminosity is directly proportional to the  radius $R$ and temperature $T$ of a star and is given by:
 $$L=\sigma 4 \pi R^2T^4$$

Inspite of the lower temperature of Betelguese, due to it's  large size,  its luminosity is 125,000 $\times$ L $_{\odot}$. Betelguese has used up the hydrogen fuel in it's core and is presently fusing helium into carbon, which is very efficient and hence generates lot of heat. In the process, it blows a very strong wind of material away from it $\approx  10^{-6}$ M$_{\odot}$  yr$^{-1}$. In comparison the Sun loses $< 10^{-12}$ M$_{\odot}$ yr$^{-1}$.

\section{Evolution of High Mass Stars}
\rightHighlight{\textbf{Types of nuclear reactions:}\\
\textbf{PP chain}\\
4H$^1 \rightarrow$ 4He + 2e+ 2$\gamma$ + 2$\nu$e  \\
\textbf{CNO Cycle}\\
4H$^1$ +2e + (CNO) $\rightarrow$ 4He + 2e+ 2$\gamma$ + 2$\nu$e +(CNO) \\
\textbf{Triple-$\alpha$}\\
3He$^4 \rightarrow$ 12C + $\gamma$
}
\begin{figure}[!t]
\caption{Plot of  logarithm of the relative energy output ($log \epsilon$) of PP, CNO and Triple-$\alpha$ fusion processes  versus logarithm of temperature $log \ T$. The dashed line shows the combined result of the PP and CNO processes within a star. For the temperature of Sun's core, the PP process is dominant, while at higher temperatures, more efficient nuclear reactions dominate, leading to shorter lifetimes of massive stars. (Image Credit:  Wikipedia)}
\label{nr} 
\vskip -12pt
\centering
\includegraphics[width=6cm, height=5 cm]{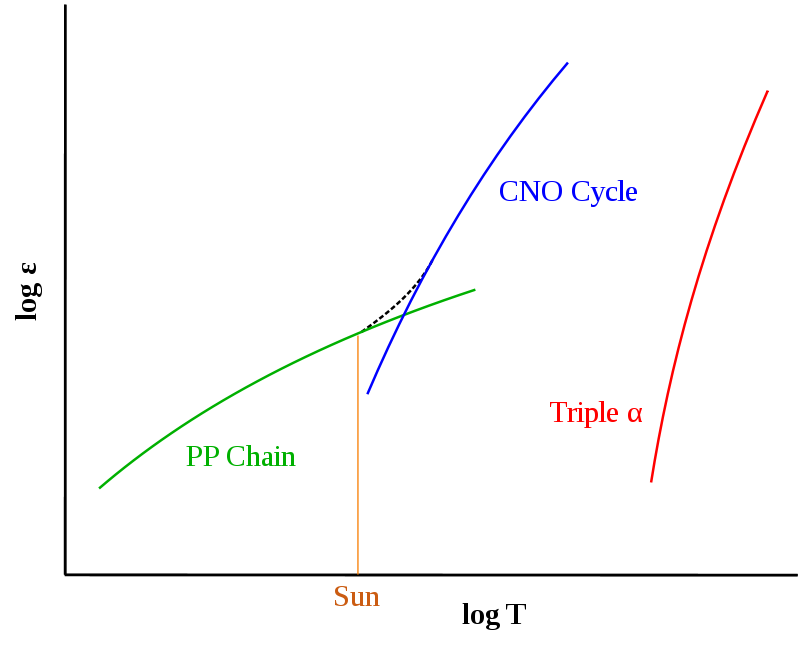}
\end{figure}

Stars basically fuse hydrogen to helium for most of their lifetimes and that phase in their life is called the main sequence. The time spent on the main sequence is given by: 
$$t_{MS} \propto M^{-2}, t_{MS} = t_{\odot} ( M_{\odot}/M)^2$$
where $ t_{\odot} =10^{10}$ yr. 

The kind of nuclear fusion reactions that take place depends on the core temperature of the star. For a less massive star, hydrogen burning is dominated by the proton--proton (PP) chain reaction which sets in  at temperatures around $4\times10^6$ K, is less efficient and ensures a longer lifetime for low mass stars. 

The CNO chain starts at approximately $15\times10^6$ K, but its energy output increases with increasing temperatures so that it becomes the dominant source of energy at approximately $17\times10^6$ K. The Sun has a core temperature of around $15.7\times10^6$ K, and only 1.7\% of helium  nuclei produced in the Sun are due to the CNO cycle.  At higher temperatures the triple-$\alpha$ reaction sets in. The CNO cycle and triple-$\alpha$ are far more efficient ensuring the star is more luminous but finishes its fuel much faster (Fig \ref{nr}). This explains the difference in lifetimes of stars with varying masses (Table. \ref{life}).

\begin{table}[h]
\caption{Lifetimes of stars as a function of mass and spectral type}
\label{life}
\centering
\vskip -12pt
\begin{tabular}{ccc}
\hline
Mass  (M$_{\odot}$)& Time (yr) & Spectral Type\\
\hline
60&	3 million	&O3\\
30&	11 million	&O7\\
10&	32 million	&B4\\
3&	370 million	&A5\\
1.5&	3 billion	&F5\\
1&	10 billion	&G2 (Sun)\\
0.1&	1000's billions	&M7\\

\hline
\end{tabular}

\vskip 12pt
\end{table}

Throughout the main sequence phase, there is a constant balance between the pressure generated by energy release and the weight of the star, called hydrostatic equilibrium and the star is stable. 

\begin{figure}[!t]
\caption{The onion-like layers of a massive, evolved star just before core collapse (not to scale) (Image Credit:  Wikipedia)}
\label{shells} 
\vskip -12pt
\centering
\includegraphics[width=5cm, height=5 cm]{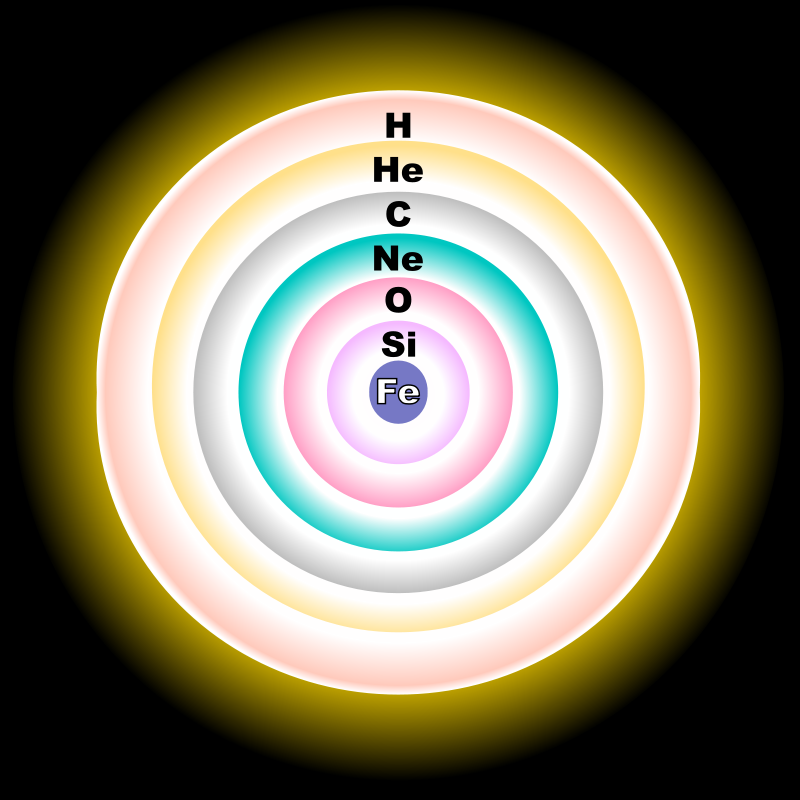}
\end{figure}

On completion of hydrogen fusion in the core, outward radiation pressure decreases. This causes gravitational contraction which in turn increases it's temperature thus making it possible for heavier elements to fuse. 
\rightHighlight{A supernova is an energetic and luminous explosion which occurs in the final  evolutionary stages of a massive star or in binary systems when one star is a white dwarf and it gets triggered into runaway nuclear fusion. The star either collapses to a neutron star or black hole, or it is completely destroyed. The peak optical luminosity of a supernova is comparable to that of an entire galaxy, before fading over several weeks or months. The Luminosity of Supernovae is a constant, and hence supernovae are used as standard candles to find distances to external galaxies independently.}
The helium core continues to grow on the red giant branch, while the radiative envelope expands. The star is no longer in thermal equilibrium, and hence it often shows variability. Betelgeuse is classified as a pulsating RSG. It physically expands and contracts as  heat generated from its core get released. Nuclear fusion takes place in onion-shell like regions of the star with the heaviest elements being fused at the core and lighter elements on the outer regions as shown in Fig. \ref{shells}.

\begin{figure}[!t]
\caption{The Crab Nebula, the shattered remnants of a star which exploded as a supernova visible in 1054 AD (Image Credit:  Wikipedia)}
\label{crab} 
\vskip -12pt
\centering
\includegraphics[width=6cm, height=6 cm]{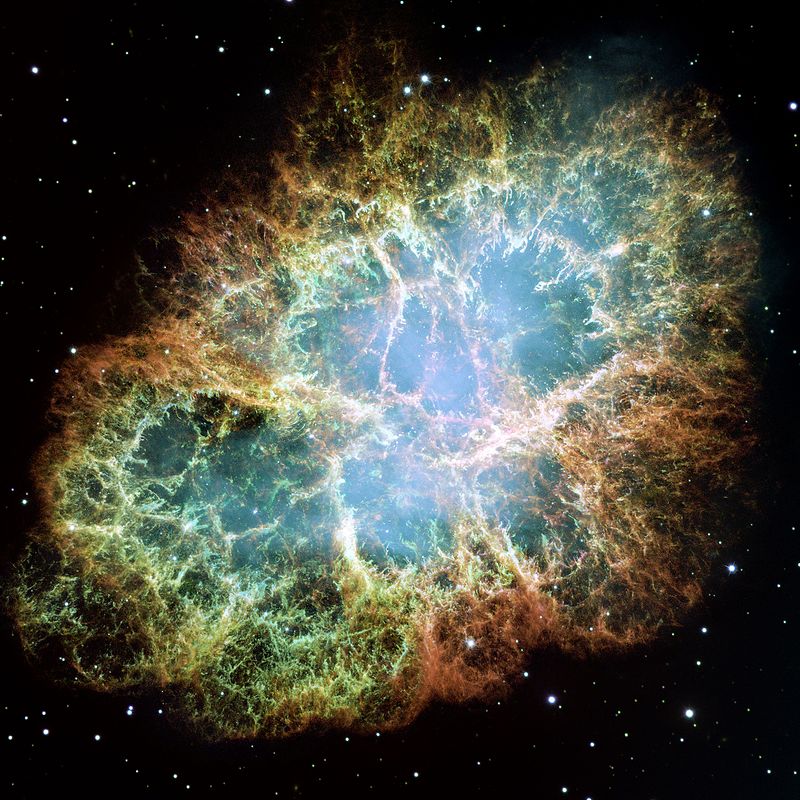}
\end{figure}

Fig \ref{crab} shows the Crab Nebula ( M1, NGC 1952, Taurus A) which is a Type II supernova remnant in the constellation of Taurus, which was recorded by Chinese astronomers in 1054 and later observed by English astronomer John Bevis in 1731. The nebula was the first astronomical object identified with a historical supernova explosion. At the center of the nebula is the Crab pulsar,  a neutron star 28--30 km across with a spin rate of 30.2 times per second, which emits pulses of radiation from gamma rays to radio waves.

\cite{dolan16} estimated that in a little less than $10^5$ yr, Betelguese  will explode as a Type II Supernova, releasing $2 \times 10^{53}$ erg in neutrinos along with $2 \times 10^{51}$ erg in explosion kinetic energy \cite{smartt2009} and leaving behind a neutron star of mass 1.5  M$_{\odot}$ \cite{dolan16}. The star would have evolved to a Si--Fe core. When this supernova explodes, it will be closer than any known supernova observed to date, and about 19 times closer than Keplerʼs supernova. Assuming that it explodes as an
average SN II, the optical luminosity will be approximately -12.4 magnitudes, becoming brighter than the full moon. The X-ray and $\gamma$--ray luminosities may be considerable, though not enough to  penetrate the Earthʼs atmosphere. The interaction of such a supernova shock with the heliosphere has been studied in detail in \cite{fields2008}. The study demonstrated that unless a typical supernova occurs within a distance of 10 pc, the bow-shock compression of the heliosphere occurs at a distance beyond 1 AU. 
\leftHighlight{The Hertzsprung--Russell diagram is a  plot of stars showing the relationship between the stars' absolute magnitudes or luminosities versus their stellar classifications or effective temperatures. It is a graphical tool that astronomers use to classify stars according to their luminosity, spectral type, color, temperature and evolutionary stage. Stars in the stable phase of hydrogen burning lie along the Main Sequence according to their mass.}
For the adopted distance of 197 pc, the passing supernova remnant shock is not likely to directly deposit material on Earth. 

\section{The Hertzspring--Russell Diagram}

The Hertzsprung--Russell diagram (HR) diagram is a scatter plot of the absolute magnitude or luminosity of stars  versus their stellar classifications or effective temperatures. The diagram was created independently around 1910 by Ejnar Hertzsprung and Henry Norris Russell and stellar evolution can be best represented by the position of a star on the diagram. All important parameters of stars, i.e., its mass, radius, temperature, luminosity, age and evolutionary stage can be found once we know the position of a star on this diagram. We also use this diagram to understand the evolution of stars.  

\begin{figure}[!t]
\caption{An HR diagram with the instability strip and its components highlighted.  (Image Credit:  Wikipedia)}
\label{hr} 
\vskip -12pt
\centering
\includegraphics[width=6cm, height=6 cm]{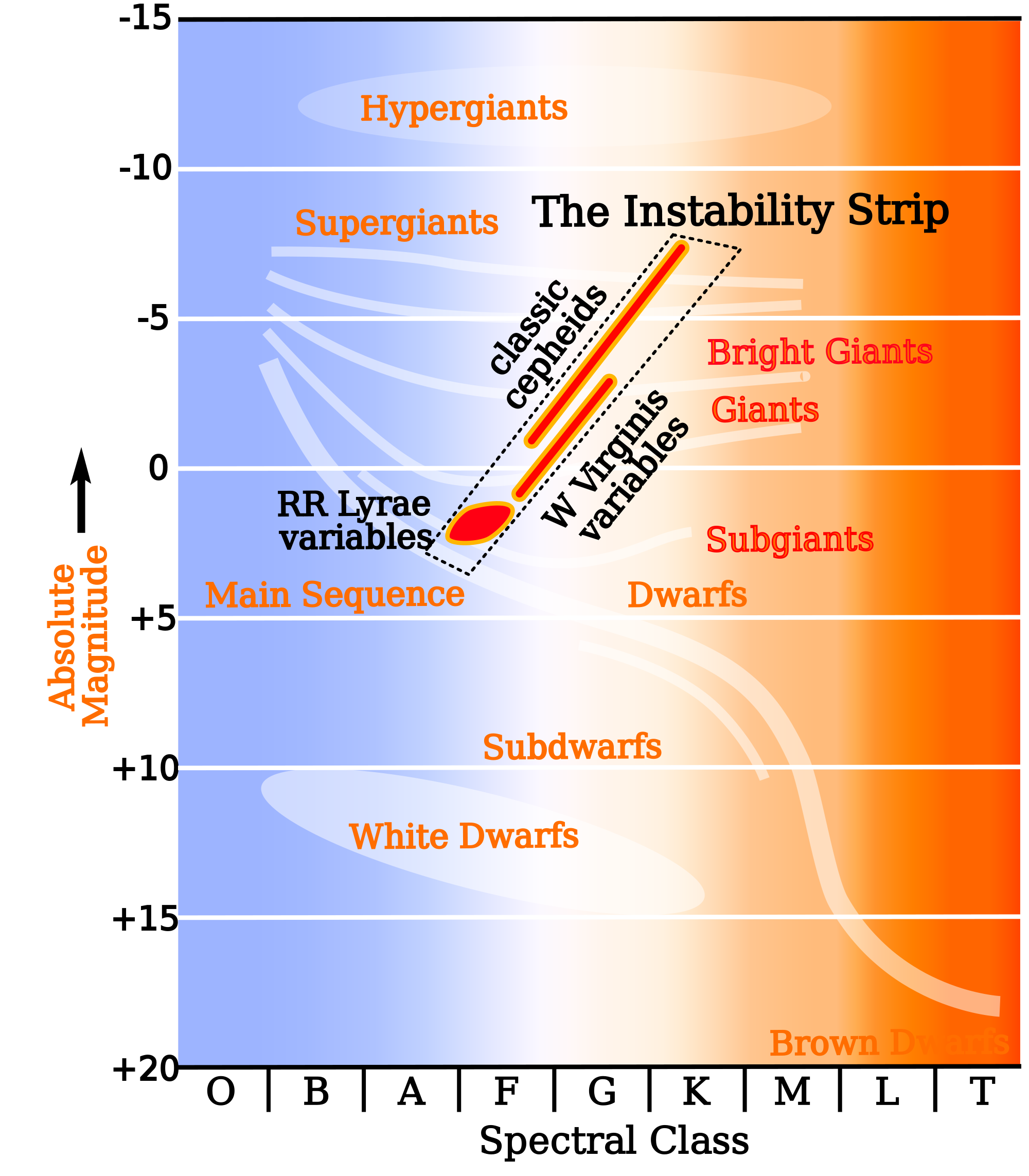}
\end{figure}

Most of a star's life is spent along the line called the main sequence. During this stage,  stars are fusing hydrogen in their cores.  After a time period in Fig \ref{life}, helium fusion begins in the core and hydrogen burning in a shell surrounding the core. The star is unstable and various categories of variables which include RR Lyraes and Cepheid variables marked in Fig.\ref{hr} can be found  in the instability strip.

\rightHighlight{The instability strip refers to an area of the Hertzsprung--Russell diagram where lie several related classes of pulsating variable stars:Delta Scuti variables, SX Phoenicis variables, and rapidly oscillating Ap stars (roAps) near the main sequence; RR Lyrae variables where it intersects the horizontal branch; and the Cepheid variables where it crosses the supergiants.}

After the completion of this red giant phase, massive stars will go supernova and the core will end up as a white dwarf, neutron star  or a black hole. Nuclear fusion for elements upto iron is exothermic and takes place in  massive stars. For all elements heavier than iron, fusion is an endothermic process and hence that can only take place when a large amount of energy is released or is available, like in a supernova.  Hence all heavy elements (heavier than iron) have their origin in such energetic processes. 

 In 1973, Carl Sagan, the renowned astronomer, published `The Cosmic Connection: An Extraterrestrial Perspective' which included the following passage `All of the rocky and metallic material we stand on, the iron in our blood, the calcium in our teeth, the carbon in our genes were produced billions of years ago in the interior of a red giant star. We are made of star-stuff'.

\section{Variability of Betelguese}

\begin{figure}[!t]
\caption{Light curve showing periodic variations in Betelgeuse's brightness from 1979 to the present using V-band photometry. The vertical axis plots magnitude, the horizontal plots the time in Julian dates.  (Image Credit:  AAVSO)}
\label{var} 
\vskip -12pt
\centering
\includegraphics[width=9cm, height=5 cm]{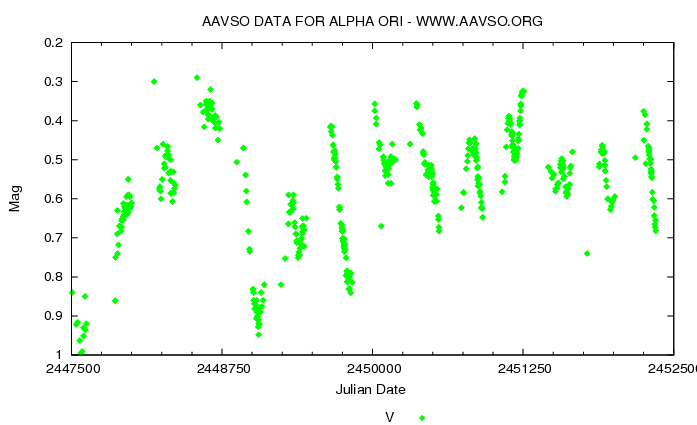}
\end{figure}
Betelgeuse is classified as a pulsating Red Super Giant (RSG). As the star is not in equilibrium, there are alternate expansions  and contractions due to imbalance between the radiation pressure created by nuclear reactions and the gravitational force. This imbalance is so great that when the star is smallest and hottest, it extends to the orbit of Mars if put in place of the Sun and when largest and coolest it can fill up Jupiter's orbit. 

\leftHighlight{A variable star is a star whose brightness fluctuates. This variation may be caused by a change in emitted light or by something partly blocking the light, so variable stars are classified as either: Intrinsic variables, whose luminosity changes due to physical processses on the star or Extrinsic variables, whose apparent changes in brightness are due to changes in the amount of their light that can reach Earth; for example, in eclipsing binaries. \\
Most stars have a variation in their luminosity, the Sun, for example, varies in its luminosity by about 0.1\% over it's 11-year solar cycle}

The variability of Betelgeuse is  semi-regular with multiple periods of variation. The primary pulsations repeat every 425 days, but the star also shows additional periods corresponding to 100-180 days and 5.9 years. The star has dark patches that are huge sunspots as well as bright regions of rising gas leading to further fluctuations in its brightness. What we know is that Betelgeuse is in a bright phase and will finally run out of fuel, collapse, and explode as a Type II supernova.

 Astronomers Edward Guinan and Richard Wasatonic (Villanova University) and amateur astronomer Thomas Calderwood, had been monitoring the star for more than 25 years. They reported a fall to magnitude 1.29 on December 20th  2019 using precise V-band photometry, this being the lowest in records. In Astronomical Telegram  No. 13365 Guinan concluded that  `The current faintness of Betelgeuse appears to arise from the coincidence of the star being near the minimum light of the 5.9~yr light-cycle as well as near the deeper than usual minimum of the 425~day period.' In effect, the star's overlapping cycles have created a sort of superminimum.

\section{Spectrophotometric Observations}

Levesque \& Massey (2020) determined the effective temperature of Betelguese to be $3600\pm25$ K by doing optical spectrophotometry of the star and comparing this spectrum to stellar atmosphere models for cool supergiants. They found that though this is slightly cooler than previous measurements  taken prior to Betelgeuse's recent lightcurve evolution, this drop is insufficient to explain Betelgeuse's recent optical dimming.
\rightHighlight{\cite{leves20} used observations were made on 2020 Feb 15 using the DeVeny low-to-moderate resolution optical spectrograph on the 4.3-meter Lowell Discovery Telescope (LDT). The wavelength coverage was 4000-6700~$\AA$ with a resolution of 8.0~$\AA$}

\begin{figure}[!t]
\caption{Betelgeuse blowing a huge amount of gas and dust into space, seen in this infrared image from the Very Large Telescope. Betelgeuse is in the very center, and so bright that the area around it has been scaled down in contrast so that the far fainter outer material can be viewed. (Image Credit: ESO/P. Kervella)}
\label{vlt} 
\vskip -12pt
\centering
\includegraphics[width=6cm, height=5 cm]{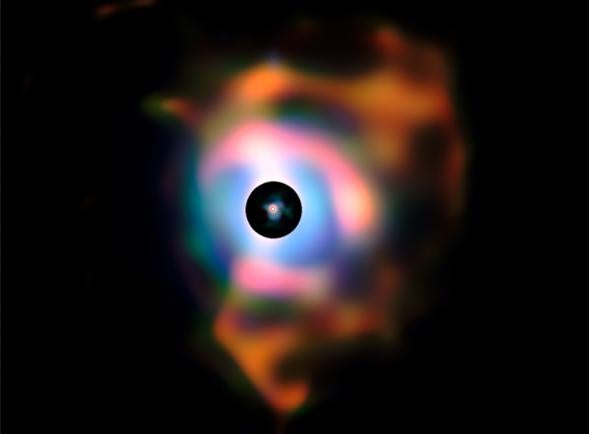}
\end{figure}

 A fall in the effective temperature from 3650 K to 3600 K  would correspond to a decrease in $V$ magnitude of only 0.17. Hence, decrease in temperature of the star cannot be the cause of the dimming. Hence, based on  spectrophotometry, it was proposed that as Betelgeuse has been showing recent episodic mass loss, large-sized dust grains could be the best explanation for the dimming of the star.  Circumstellar dust composed of larger grains would cause extinction or loss of light more in the optical part of the spectrum compared to the  bluer longer wavelengths, causing a dimming of the $V$ magnitude of Betelgeuse.  This explanation agrees with the lack of significant changes seen in the star's $T_{eff}$ as well as with the  consistent amount of extinction due to dust towards Betelgeuse combined with an increase in flux at the bluest wavelengths in the 2020 spectrum. The Fig. \ref{vlt} shows VLT image of Betelguese showing the huge amount of gas and dust around the star, thus supporting this possible explanation.

\section{Conclusion}
The present dimming of  Betelguese can be explained by episodic mass loss that let out large-size grain particles that obscures the light from Betelguese making it appear dimmer. So probably too early for supernova...

Recent observations show that Betelguese has been brightening up since 20 th of February 2020.  Multi-wavelength observations, particularly in the ultraviolet and infrared, are needed to understand and evaluate the stellar processes taking place in Betelgeuse to completely understand the variation in luminosity of the star.
\begin{figure}[!t]
\caption{Since 20 Feb 2020, Betelgeuse appears to be brightening again. (Image Credit:  AAVSO)}
\label{sphot} 
\vskip -12pt
\centering
\includegraphics[width=9cm, height=5 cm]{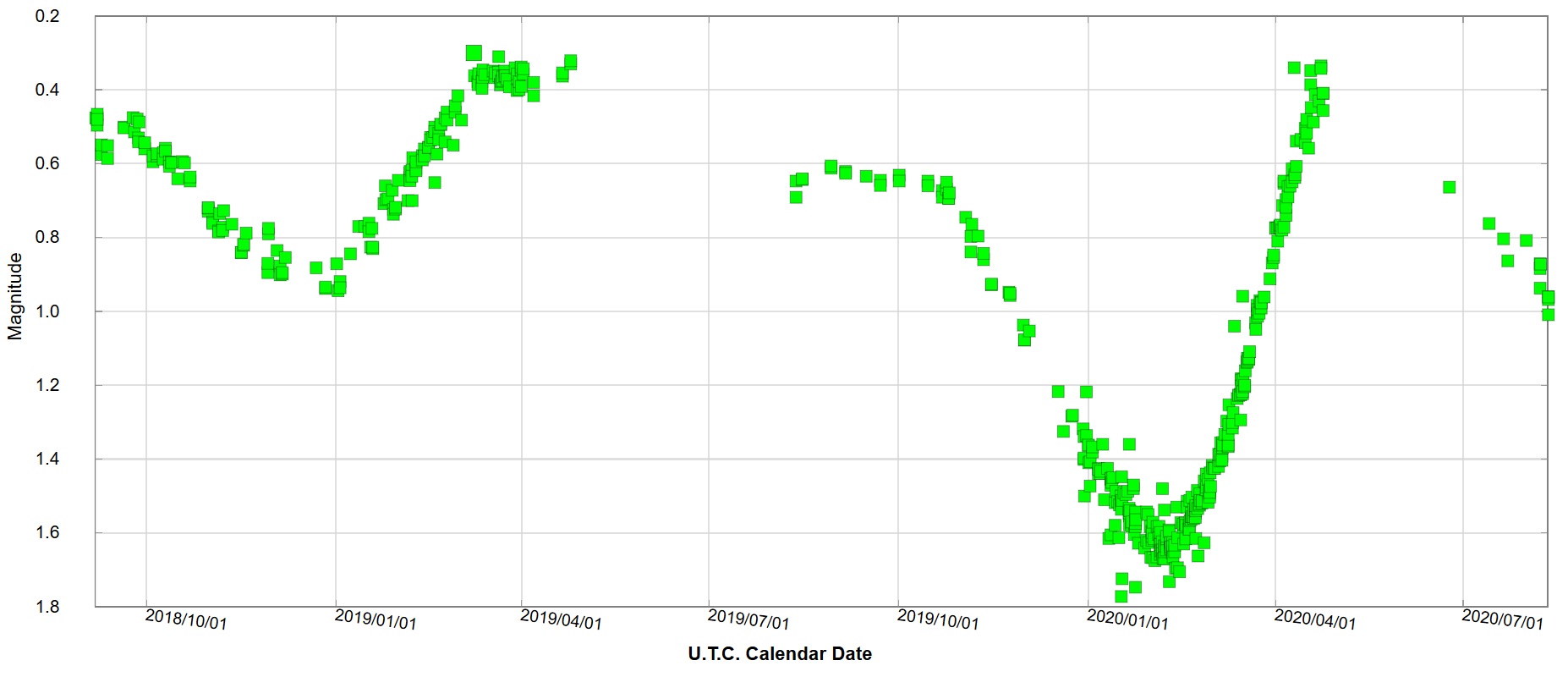}
\end{figure}

\leftHighlight{Fields et al. (2008) demonstrated that unless a typical supernova occurs within a distance of 10 pc, the bow-shock compression of the heliosphere occurs at a distance beyond 1 AU. Hence supernovae beyond 10 pc are safe for us!!!!}  

Eventually Betelguese will go supernova !!! At that distance,  it'll be as bright as the full Moon! Luckily, too far away to hurt us. And that too after 100,000 yrs! Octillions ($10^{27}$) of tons of matter would be zooming into space in all directions at a fraction of the speed of light!  The shock wave will take $6\times 10^6$ yrs to reach us, at 13~kms$^{-1}$.  The shock  will stop  outside the Earth's orbit.
We're safe.
 
But poor Orion..... with Betelgeuse gone,  he'll be missing his right shoulder that we are all so familiar with!

But till then, the star closest to going supernova is Eta Carinae which is is a two-star system about 10 times further away that's already showing signs for decades now and it will light up our skies in the daytime. Keep watching!!!!

\section*{Acknowledgements}
The author would like to thank the referee for her/his valuable comments that helped improve the content of the article.


\end{document}